\documentclass[twocolumn,a4paper,showpacs]{revtex4}
\usepackage{graphicx}
\begin{document}
\title{Transport through a Strongly Correlated Quantum-Dot with Fano Interference}
\author{B. H. Wu$^{(1,2)}$, J. C. Cao$^{(2)}$ and Kang-Hun Ahn$^{(1)}$}
\affiliation{(1) Department of Physics, Chungnam National University, Daejeon 305-764, Republic of Korea\\
(2) State Key Laboratory of Functional Materials for Informatics,
Shanghai Institute of Microsystem and Information Technology,
Chinese Academy of Sciences, 865 Changning Road, Shanghai 200050, People¡¯s Republic of China}
\begin{abstract}
We present the transport properties of a strongly correlated quantum dot attached
to two leads with a side coupled non-interacting quantum dot.
Transport properties are analyzed using the slave boson mean field theory which
is reliable in the zero temperature and low bias regime. It is found that the
transport properties are determined by the interplay of two fundamental physical phenomena,i.e. the Kondo
effects and the Fano interference. The linear
conductance will depart from the unitary limit and the zero bias anomaly will be suppressed in the presence of
interdot coupling.
The zero bias shot noise Fano factor increases with the interdot coupling and tends to the Poisson value.
The shot noise Fano factor shows a non-monotonic behavior as
a function of the interdot coupling for various side dot energy levels.
\end{abstract}
\date{\today}
\maketitle
\section{introduction}
In the past years, Kondo effect in artificial impurities has
attracted much attention.
The Kondo effect\cite{Hewson} in QD originates from the strong many-body correlations
due to the internal spin degrees of freedom
 when the temperature $T$ is lower than the characteristic Kondo temperature.
At equilibrium, a resonant peak of the density of states forms at the Fermi energy of the leads due to
Kondo effect.
Such effect results in the unitary limit of the linear conductance
at very low temperature and the zero bias anomaly of the differential conductance\cite{PRL611768,
PRL673720,PRL702601}.

By coupling two QDs, one gets a double quantum dot (DQD) \cite{RMP751}.
The interplay of the correlation and quantum interference
effects due to different DQD arrangement with respect to the leads provides us useful model systems for the
study of mesoscopic physics.
Comparing with the usual series and parallel strongly correlated DQD
\cite{PRL823508,Sci2932221,PRL851946,PRB63125327,PRB66125315,PRB69235305},
study on the T-shaped DQD is relatively insufficient\cite{PRB63245326,PRB71075305}.
In the T-shaped DQD, only one central quantum dot
connects to the two leads, while the other one is side coupled to the central dot. In such configurations,
there exist two pathes when an electron transport through
the DQD. One is the directly transport through the central QD. The other path includes the side QD.
Thus, Fano interference occurs. Fano interference has been proven to be a very interesting phenomena
for the study of quantum coherence in QDs
\cite{PRB67195335,PRL865128,PRL87156803,PRL93106803,PRB63113304,PRB65085302,PRB69165325,JPCM168285}.
The interplay of the correlations in the QDs and the
Fano interference is expected to provide interesting phenomena.
By coupling a side dot to an ideal quantum wire,
the conductance was found to be suppressed
due to the formation of the Kondo resonance at the Fermi energy on the side QD\cite{PRB63113304,PRB65085302}.
Equilibrium properties of the T-shaped
DQD was very recently numerically investigated by using the numerical renormalization group (NRG) method
\cite{PRB71075305}.
The NRG technique is known to give  reliable results on equilibrium systems but not valid in
describing the nonequilibrium situations. Kim $et\ al.$ investigated the nonequilibrium transport properties
of the T-shaped DQD
by using the nonequilibrium Green function combined with the noncrossing approximation\cite{PRB63245326}.
In their model,
only the side dot was assumed to be strongly correlated while the central dot was noninteracting.
The Kondo peak of density of states at the Fermi energy was not formed at the central dot. As a consequence,
the zero bias anomaly was not seen in their results.

\begin{figure}
\includegraphics[width=7cm]{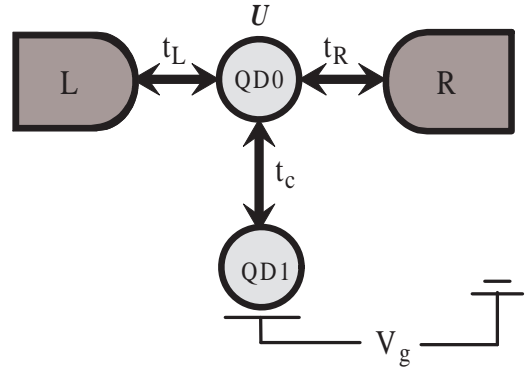}
\caption{Schematic plot of the nanodevice. Only one quantum dot QD0 is connected with
both leads. Another noninteracting quantum dot QD1 is side coupled to QD0. Energy level of
QD1 can be adjusted via the gate voltage or a magnetic field.
}\label{Fig1}
\end{figure}

In this paper, we systematically exploring the nonequilibrium transport
properties of a T-shaped DQD where the central dot operates
in the Kondo regime. The side dot is assumed to be noninteracting so that
the Fano interference can be readily adjusted by changing interdot coupling and
the energy level of the side dot.
Current and shot noise properties of the DQD are investigated.
Both Kondo effect and Fano interference play  important  roles in determining
the transport properties of the T-shaped DQD device.

Our results were obtained by employing the slave boson mean field theory (SBMFT)\cite{{PRB293035,PRB355072}}
with the use of nonequilibrium Green function following \cite{PRL851946}.
The SBMFT is numerically simple and efficient for the study of
transport properties in strongly correlated QDs. As a strong coupling approach, the SBMFT has been proven to be
reliable for the low voltage at zero temperature.
In our case, the SBMFT enables us to include nonperturbatively the interdot coupling,
so that the interplay of the correlation and the interference can be explicitly studied.
Due to the discrete nature of charge particles, measurement of the shot noise can reveal
the correlations between carriers out of equilibrium\cite{Nature392658,PR3361}.
Unfortunately, it is usually nontrivial to obtain an expression for the shot noise of strongly correlated systems
\cite{PRB71035315,PRL88116802}.
However, such an expression is straightforward
within the mean field approximation where one can apply the Wick theorem \cite{PRB71035315,JPCM144963}.
Despite these merits, the SBMFT has its own limitations.
It takes into account spin fluctuations but neglects
the charge fluctuations. Therefore, calculations based on the SBMFT are restricted to the low bias limit.
Otherwise, one has to rely on more advanced approaches such as the noncrossing approximation.

\section{model and formulation}

We consider nonequilibrium transport properties of a T-shaped DQD device at zero temperature.
The schematic diagram of our device is presented in Fig. \ref{Fig1}.
The central part of our device has two QDs.
A single energy level is included on each dot.
In our configuration, only one dot QD0 is connected to both leads via tunnelling coupling $t_\alpha$ ($\alpha=L,R$).
There is a strong intradot Coulomb interaction $U$ on QD0. Another noninteracting dot QD1 is side coupled to
QD0 via interdot tunnelling coupling $t_c$. The energy level of QD0 $\epsilon_{0\sigma}$ is far below the Fermi energy
while the energy level of QD1 $\epsilon_{1\sigma}$ can be tuned by the external gate voltage.
Here, $\sigma$ represents spin.
In the following discussion, $U$ is sufficiently large ($U\rightarrow\infty$),
so that double occupancy of QD0 is forbidden.
The infinite $U$ slave boson language was adopted to describe the strongly correlated QD.

The Hamiltonian of our model reads:
\begin{eqnarray}
H&=&\sum_{k\alpha,\sigma}\epsilon_{k\alpha,\sigma}c^\dag_{k\alpha,\sigma}c_{k\alpha,\sigma}
+\sum_{\sigma}\epsilon_{0\sigma}f^\dag_{\sigma}f_\sigma+\sum_{\sigma}\epsilon_{1\sigma}d^\dag_{\sigma}d_{\sigma}
\\ \nonumber & &
+{{1}\over{\sqrt{N}}}\sum_{k\alpha,\sigma}t_{\alpha}(c^\dag_{k\alpha,\sigma}b^\dag f_{\sigma}+H.C.)\\ \nonumber
& &+{t_c\over\sqrt{N}}\sum_{\sigma}(d_{\sigma}^\dag b^\dag f_\sigma+H.C.)\\ \nonumber
& &+\lambda(b^\dag b+\sum_{\sigma}f^\dag_{\sigma}f_{\sigma}-1),
\end{eqnarray}
where $c^\dag_{k\alpha,\sigma}(c_{k\alpha,\sigma})$ is the creation (annihilation) operator for an electron with
spin $\sigma$ and quantum number $k$ in the $\alpha$ lead.
In the slave boson representation, the creation (annihilation)
operator for an electron in the $N=2$ fold degenerate QD0 with spin $\sigma$ is
replaced by $f_\sigma^\dag b\ (f_\sigma b^\dag)$ where
$f_\sigma$ ($f^\dag_\sigma$) is the pseudofermion operator that annihilates (creates) a singly occupied state in QD0
and  $b$ ($b^\dag$) is the slave boson operator which annihilates (creates) an empty state in QD0.
The first line of the Hamiltonian represents electrons in the leads and the two dots. The second and third line
of the Hamiltonian describes the QD0-leads tunnelling processes and the interdot coupling.
The last term in the Hamiltonian with a Lagrange multiplier $\lambda$ represents the constraint
$\sum_\sigma f^\dag_\sigma f_\sigma+b^\dag b=1$ for QD0 to prevent the double
occupancy in the limit $U\rightarrow\infty$.

We solved this Hamiltonian by using the mean field approximation, where the boson operator is
approximated by its expectation value. A real parameter $\tilde b$
is introduced to replace the boson operators in the Hamiltonian, namely, $b(t)/\sqrt{N}=\tilde b$. This
approximation is exact for describing the spin fluctuation (Kondo regime)
in the limit $N\rightarrow\infty$ and $T=0$. By defining the renormalization parameters $\tilde t_\alpha=t_\alpha
\tilde b$, $\tilde t_c=t_c\tilde b$ and $\tilde \epsilon_{0\sigma}=\epsilon_{0\sigma}+\lambda$, the Hamiltonian
(1) is formally reduced to the free electron model:
\begin{eqnarray}
\tilde H&=&\sum_{k\alpha,\sigma}\epsilon_{k\alpha,\sigma}c^\dag_{k\alpha,\sigma}c_{k\alpha,\sigma}
+\sum_{\sigma}\tilde\epsilon_{0\sigma}f^\dag_{\sigma}f_\sigma+\sum_{\sigma}\epsilon_{1\sigma}d^\dag_{\sigma}d_{\sigma}
\\ \nonumber & &
+\sum_{k\alpha,\sigma}\tilde t_{\alpha}(c^\dag_{k\alpha,\sigma} f_{\sigma}+H.C.)\\ \nonumber
& &+{\tilde t_c}\sum_{\sigma}(d_{\sigma}^\dag  f_\sigma+H.C.)\\ \nonumber
& &+\lambda(N \tilde b^2 -1).
\end{eqnarray}

Now, the task is to self-consistently determine the set of parameters: $\tilde b$ and $\lambda$.
Following the Ref. \cite{PRL851946} one can first write down the constraint and the equation of motion of the slave boson
operators. Thus we have two equations with the above two unknowns. Next, the two equations are closed
with the nonequilibrium Green's functions of the central dot by
applying the equation of motion and the analytic continuation rules \cite{HaugJauho}
of the lead-dot lesser Green function.
These equations in the Fourier space are given by
\begin{equation}\label{CEQ1}
\tilde b^2-i\sum_\sigma\int {d\epsilon\over{4\pi}}G^<_{00,\sigma}(\epsilon)={1\over 2},
\end{equation}
\begin{equation}\label{CEQ2}
\lambda \tilde b^2=i\sum_\sigma\int{d\epsilon\over 4\pi}(\epsilon-\tilde\epsilon_{0,\sigma})G^<_{00,\sigma}(\epsilon),
\end{equation}
where $G^<_{00,\sigma}(\epsilon)=i<f_\sigma^\dag f_\sigma>$ is the lesser Green function of QD0.
The retarded and the lesser Green function of QD0 are frequently used in the following calculations.
These functions can be obtained by the equation of motion approach and they are given by
\begin{equation}
G_{00,\sigma}^r(\epsilon)={\epsilon-\epsilon_{1\sigma}\over
(\epsilon-\tilde\epsilon_{0\sigma}-i\tilde\Gamma)(\epsilon-\epsilon_{1\sigma})-\tilde t_c^2},
\end{equation}
and
\begin{equation}\label{LESSG}
G_{00,\sigma}^<(\epsilon)={2i(\tilde\Gamma_Lf_L(\epsilon)+\tilde\Gamma_Rf_R(\epsilon))(\epsilon-\epsilon_{1\sigma})^2
\over\mid(\epsilon-\tilde\epsilon_{0\sigma}-i\tilde\Gamma)(\epsilon-\epsilon_{1\sigma})-\tilde t_c^2\mid^2},
\end{equation}
where $f_\alpha$ is the Fermi distribution function in lead $\alpha$. 
$\tilde\Gamma_\alpha=\tilde b^2\Gamma_\alpha$ and $\tilde\Gamma=\tilde b^2\Gamma=\tilde\Gamma_L+\tilde\Gamma_R$
are the renormalized linewidth functions.
The linewidth function is defined as $\Gamma_\alpha(\epsilon)=\pi\sum_{k\alpha,\sigma}t_\alpha^2\delta(\epsilon-
\epsilon_{k\alpha,\sigma})$.
In the wide band limit, $\Gamma_\alpha$ can be reduced to an energy-independent constant for $|\epsilon|<D$
($D$ is the band width).

The two parameters ($\tilde b$ and $\lambda$) can now be self-consistently obtained by
solving Eqs. (3) and (4) with the help of Eq. (6). When the side
dot is decoupled to QD0, the Kondo temperature of QD0
($\epsilon_0=\epsilon_{0,\sigma}$) at equilibrium
is given by $T_K^0=D e^{-\pi|\epsilon_0-E_F|/\Gamma}$.
The self-consistently determined parameters
$\tilde \epsilon_{0,\sigma}$ and $\tilde\Gamma_\alpha$ give exactly the position and the width of
the Kondo peak in QD0.

After obtaining these parameters self-consistently, the transmission probability $T_\sigma(\epsilon,V_{LR})$
can be found from
\begin{equation}
T_\sigma(\epsilon,V_{LR})=G_{00,\sigma}^a\tilde\Gamma_LG^r_{00,\sigma}\tilde\Gamma_R,
\end{equation}
where $V_{LR}$ is the small dc voltage applied across QD0 and  the advanced Green function $G^a$
can be calculated from the Hermite conjugate of $G^r$.
It should be noted that comparing with the noninteracting cases, the transmission probability
is a function of $V_{LR}$ due to the self-consistently determined renormalized parameters.

We studied the current and shot noise properties of the T-shaped DQD device. The current is defined as the change of
electron numbers in the leads. Current noise is caused by the time-dependent fluctuation. At $T=0$, only
the shot noise is nonzero while the thermal noise is fully suppressed.
In deriving the shot noise expression for systems with many body interaction, a direct application of
Wick theorem is not permitted due to the existence of quartic terms in the Hamiltonian. This obstacle
can be circumvented by first formally reducing
 the quartic Hamiltonian to the quadratic form under the mean field approximation. Then, an application of
 the Wick theorem is straightforward.
After some algebra, expressions for the current and zero frequency shot noise spectrum
are given by\cite{PRL682512, PRB69235305}
\begin{equation}
I={e\over h}\sum_\sigma\int_0^{eV_{LR}} d\epsilon T_\sigma(\epsilon,V_{LR}),
\end{equation}
and
\begin{equation}
S={2e^2\over h}\sum_\sigma\int_0^{eV_{LR}}d\epsilon T_\sigma(\epsilon,V_{LR})(1-T_\sigma(\epsilon,V_{LR})),
\end{equation}
where $e$ and $h$ are the charge unit and Planck constant. These results are obtained at zero temperature where
the Fermi distribution function becomes a unit step function.
The shot noise Fano factor $\gamma=S/2eI$ is introduced to characterize the deviation of the shot noise from the
Poisson value of shot noise $2eI$.

\section{numerical results and discussion}
In this section, we discuss the transport properties of the T-shaped DQD at zero temperature. In the following,
$\Gamma$ is taken as the energy unit and $E_F$ at equilibrium is set to be zero
as energy reference. The energy level of the central dot
is fixed at $-3.5$ and the conduction band has a constant DOS with the band width $D = 60$.
At equilibrium and without interdot coupling, the Kondo temperature $T^0_K$ of  QD0  is
then approximate to be $10^{-3}$.
The energy level of QD1 $\epsilon_1$ and the interdot coupling strength $t_c$ are assumed to be readily adjustable.

\begin{figure}
\includegraphics[width=8cm]{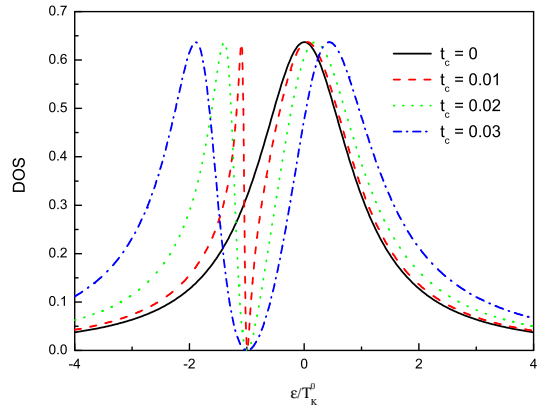}
\caption{Density of state at QD0 $\rho_0$ for different coupling strength $t_c$.
The energy level of QD1 is fixed at $-T_K^0$.}\label{DOS}
\end{figure}

\begin{figure}
\includegraphics[width=8cm]{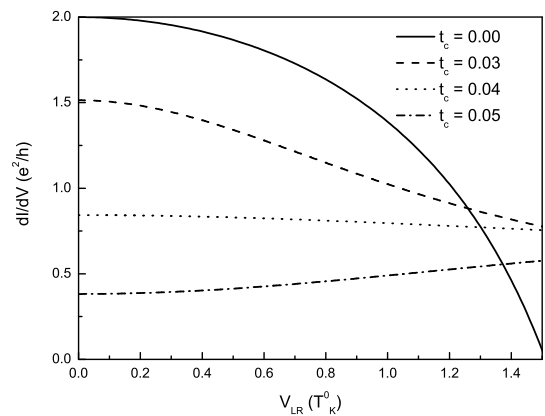}
\caption{Differential conductance as a function of the applied voltage for
various interdot couplings $t_c=0, 0.03, 0.04$ and $0.05$. The energy level of QD1
is fixed at $T^0_K$.}\label{Fig2}
\end{figure}

Let's first calculate the local density of states (DOS) in QD0 which is given by
\begin{equation}
\rho_0(\epsilon)=-{1\over\pi}\tilde b^2\sum_\sigma{\rm Im}G_{00,\sigma}^r(\epsilon),
\end{equation}
where the retarded Green function is given in Eq.(5).
Fig.\ref{DOS} presents how the calculated local DOS at QD0 changes with the interdot coupling strength $t_c$.
The side energy level is fixed at $-T_K^0$.
The DOS without Fano interference ($t_c=0$) is presented for comparison.
For $t_c=0$, the DOS shows a Breit-Wigner line shape centered at the Fermi energy. This DOS peak is solely
determined by the Kondo effects.
For $t_c\neq 0$, The Fano interference begins playing an role in complicating the DOS at QD0.
The DOS profile with Fano interference presented in Fig.\ref{DOS} can
be decomposed into a Breit-Wigner and a Fano line shape.
Inserting Eq. (5) into Eq. (10) and taking into account the spin degeneracy
($\epsilon_{0\sigma}=\epsilon_0,\epsilon_{1\sigma}=\epsilon_1$), the DOS expression can be written as
\begin{equation}
\rho_{0}(\epsilon)={2\tilde b^2\over \pi}{\tilde\Gamma^2(\epsilon-\epsilon_1)^2\over
(\epsilon-\epsilon_+)^2(\epsilon-\epsilon_-)^2+\tilde\Gamma^2(\epsilon-\epsilon_1)^2},
\end{equation}
where $\epsilon_{\pm}={\tilde\epsilon_0+\epsilon_1\over 2}\pm\sqrt{\tilde t_c^2+({\tilde\epsilon_0-\epsilon_1\over 2})^2
}$.
Eq.(11) can readily be approximated by a sum of a Breit-Wigner function and a
Fano function where the details of these two functions are not presented here for simplicity.
One can see that the DOS profile has two maxima at $\epsilon=\epsilon_{\pm}$. Between these two peaks, the DOS
dived to zero at $\epsilon=\epsilon_1$.
The distance between the two DOS maxima is $\epsilon_+-\epsilon_-=2
\sqrt{\tilde t_c^2+({\tilde\epsilon_0-\epsilon_1\over 2})^2}$.
The distance between the two peaks is a complex function of
all the parameters of the DQD device which must be determined self-consistently.
With increasing the interdot coupling $\tilde t_c$, the two peaks move to opposite directions
while the zero DOS remains unchanged at the side dot energy level.
When $\tilde t_c$ goes down, this distance gets smaller and the Fano peak becomes sharper.
An extreme case is that as $\tilde t_c$ reaches zero, the Fano resonance width becomes zero.
The Fano peak vanishes and the DOS profile reduces to the simple Breit-Wigner function.

In Fig.\ref{Fig2}, we plot the nonlinear differential conductance $dI/dV$ as a function of the applied voltage
for various interdot couplings. The differential conductance is obtained by numerically differentiating
Eq. (8) with respect to the bias voltage. For $t_c=0$, only one dot (QD0) is active.
The unitary limit of the conductance at zero bias is reached due to the Kondo resonance.
The differential conductance shows a zero bias anomaly which is the main feature of the differential conductance
in the Kondo regime. When $t_c$ is nonzero, the differential conductance line shape is modified.
The zero bias conductance departs from the unitary limit and the
zero bias anomaly peak is suppressed with increasing interdot coupling
due to the interplay of the Fano interference and the Kondo effect.  When $t_c$ is strong enough, the zero bias
anomaly eventually vanishes.

\begin{figure}
\includegraphics[width=8cm]{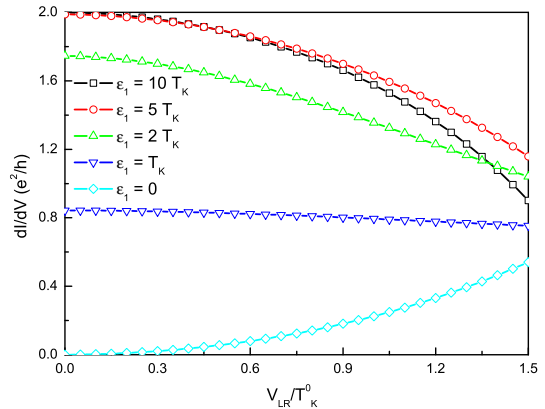}
\caption{Differential conductance with different side dot energy levels. The
interdot coupling is fixed at $t_c=0.04\Gamma$.}\label{Fig3}
\end{figure}

The Fano interference is dependent on not only the interdot coupling but also the
side dot energy level.
In Fig.\ref{Fig3}, the differential conductance as a function of the applied voltage is presented
for different side dot energy level position ($\epsilon_1$). The interdot coupling strength is fixed at $t_c=0.04$.
When $\epsilon_1$ is near the Fermi energy, the Fano interference plays an important
role in weakening the Kondo resonance. One can see from Fig.\ref{Fig3} that the zero bias anomaly
for $\epsilon_1=0$ and $\epsilon_1=T^0_K$ can not survive in the parameters given above.
The conductance at the zero bias is greatly reduced from the unitary limit. When $\epsilon_1$
is aligned with the Fermi energy ($\epsilon_1=0$), the conductance is zero due to the destructive interference
between the Kondo resonance and the side dot level.
For the case where $\epsilon_1$ is much higher than the Fermi energy, the Fano interference is expected to have a
smaller impact on the Kondo resonance. For $\epsilon_1=2 T^0_K$, $\epsilon_1=5 T^0_K$ and $\epsilon_1=10 T^0_K$,
the zero bias anomaly reappears. The zero bias conductance approaches to the unitary limit.

\begin{figure}
\includegraphics[width=8cm]{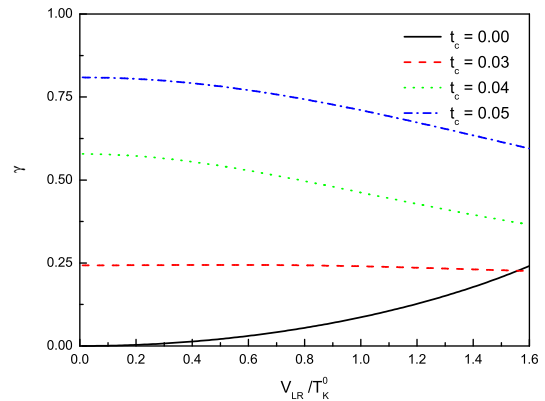}
\caption{Bias dependence of the shot noise Fano factor $\gamma$ with different interdot coupling parameters.
The side dot energy level is fixed at $\epsilon_1=T^0_K$.
}\label{Fig4}
\end{figure}

Now, we turn to the shot noise properties of the T-shaped DQD.
In Fig.\ref{Fig4}, the bias dependent Fano factor is displayed
for different interdot couplings.  The energy level of QD1 is fixed at $T_K^0$.
We are interested in the zero bias limit Fano factor which
is given by $\gamma\mid_{V_{LR}\rightarrow0}=1-T(E_F,0)$ according to Eq. (8) and (9).
For $t_c=0$, QD0 acts as a perfect transparent scatterer
dominated by the Kondo resonance. One can see from Fig.\ref{Fig4} that the Kondo unitary limit leads to
a complete suppression of the zero bias limit shot noise Fano factor.
 When the Fano interference plays its role, the Kondo resonance peak of DOS changes from a Breit-Wigner line shape to
a composition of a Breit-Wigner and a Fano line shape as can be seen from Fig.\ref{DOS}. At the same time,
the zero bias conductance will depart from the unitary limit and the
zero bias limit shot noise Fano factor will take on a nonzero value.
This enhancement of the shot noise Fano factor can be interpreted as
a result of the weakening of the DOS Kondo resonance peak at Fermi energy
by the Fano interference between the two pathways through the DQD device.
When $t_c$ becomes much stronger, The Kondo resonance at Fermi energy is destroyed by the Fano interference.
The transmission probability at $E_F$ becomes very low. As a consequence, $\gamma$ increases
very quickly and the shot noise tends to the Poisson value.

\begin{figure}
\includegraphics[width=8cm]{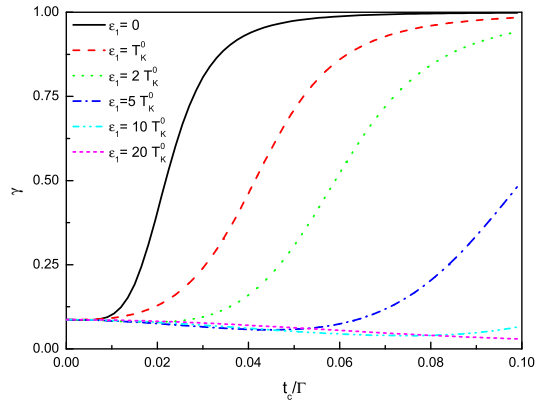}
\caption{Fano factor as a function of the interdot coupling strength for different
$\epsilon_1$. The bias is fixed at $V_{LR}=T_K^0$.}
\label{Fig5}
\end{figure}

In order to check how sensitive the shot noise is to the presence of the Fano interference,
we plot in  Fig.\ref{Fig5} the shot noise Fano factor $\gamma$ as
a function of $t_c$ at fixed  $V_{LR}$ ($V_{LR}=T^0_K$) for different $\epsilon_1$.
It is shown in Fig.\ref{Fig5} that the shot noise Fano factor with $\epsilon_1$ near $E_F$
is more sensitive to the interdot coupling than that where $\epsilon_1$ is far from $E_F$.
For cases where $\epsilon_1$ is near the Fermi energy
($\mid\epsilon_1-E_F\mid<T_K^0$), the Kondo resonance peak at Fermi energy will be
abruptly quenched by the emergence of the Fano interference. This leads to a sudden increase of $\gamma$
from a small value at zero interdot coupling to the Poisson value. Therefore, the Fano factor
$\gamma$ is  highly sensitive to the interdot coupling
when $\epsilon_1$ is near the Fermi energy as shown in Fig.\ref{Fig5}.
On the contrary, for cases where $\epsilon_1$ is far from the Fermi energy ($\mid\epsilon_1-E_F\mid\gg T_K^0$),
$\gamma$ will be less sensitive to the interdot coupling strength. This is because the presence of interdot coupling
is expected to have a weaker destructive interference on the Kondo resonance peak of DOS at $E_F$.
On can see from Fig.\ref{Fig5}, for a larger $\epsilon_1$, $\gamma$ keeps to be
small over a wider range of the tunable interdot coupling before it rises to the Poisson limit.
One may also found from Fig.\ref{Fig5} that $\gamma$ will show a non-monotonic relation with $t_c$ at finite bias.
It is interesting that the minimum of the shot noise Fano factor is not obtained at the zero interdot coupling where
the Kondo resonance dominates the transport properties.
This can be best seen when $\mid\epsilon_1\mid\gg T_K^0$
where $\gamma$ first decreases with
increasing $t_c$ when $t_c$ is below a critical value and then increase quickly with $t_c$ to reach the Poisson value.
For a larger $\epsilon_1$, one can see from Fig. 6 that this value moves to the strong interacting end.
This critical value is a function of the side dot energy level and the applied bias.
It is nontrivial to obtain an analytical expression for
the critical value because all the parameters in the Kondo regime have
to be determined self-consistently. This non-monotonic behavior of $\gamma$ is  due to the interplay of the
Fano interference and the Kondo effects. The critical value stands for the situation where the coherent transport
is optimized as the competition of the Kondo effects and Fano interference.

\section{summary}
In this work, we presented the transport properties of a
T-shaped DQD device where the central dot is in the Kondo regime.
We solved this problem by using the SBMFT at low bias voltage and zero temperature.
When the side dot is decoupled, the Kondo resonance in the central dot results in the
unitary limit of the linear conductance and the zero bias anomaly. The shot noise Fano factor at zero bias
is fully suppressed to zero.
It is more interesting to study the case where a side dot can interact with the central dot.
When an electron is emitted from one lead, there is quantum interference
between two electron waves, one directly passing through the central dot
and one traveling through the two dots, before the electron arrives at the other lead.
This Fano interference can act to weaken the
Kondo resonance at the Fermi energy.
The DOS is composed of Breit-Wigner and Fano line shapes.
As a consequence, the linear
conductance will depart from the unitary limit and the zero bias anomaly will be suppressed in the presence of
interdot coupling. The zero bias anomaly will eventually vanishes at large interdot coupling strength
or the side dot energy is tuned near the Kondo resonance peak.
The zero bias shot noise Fano factor increases with the interdot coupling and tends to the Poisson value.
Detailed study indicates that at finite bias, the shot noise Fano factor shows a non-monotonic behavior with the
interdot coupling strength as a result of the interplay of the Kondo effect and the Fano interference.

\begin{acknowledgements}
This work was supported by Korea Research Foundation, Grant
No.(KRF-2003-070-C00020).

\end{acknowledgements}



\end{document}